\documentclass[conference]{IEEEtran}
\IEEEoverridecommandlockouts
\usepackage{cite}
\usepackage{amsmath,amssymb,amsfonts}
\usepackage{algorithmic}
\usepackage{color}
\usepackage{subfigure}
\usepackage{hyperref}
\usepackage{array}
\usepackage{siunitx}
\usepackage{graphicx}
\usepackage{textcomp}
\usepackage{xcolor}
\def\BibTeX{{\rm B\kern-.05em{\sc i\kern-.025em b}\kern-.08em
    T\kern-.1667em\lower.7ex\hbox{E}\kern-.125emX}}
    
\usepackage{tikz,xcolor,hyperref}

\definecolor{lime}{HTML}{A6CE39}
\DeclareRobustCommand{\orcidicon}{%
	\begin{tikzpicture}
	\draw[lime, fill=lime] (0,0) 
	circle [radius=0.16] 
	node[white] {{\fontfamily{qag}\selectfont \tiny ID}};
	\draw[white, fill=white] (-0.0625,0.095) 
	circle [radius=0.007];
	\end{tikzpicture}
	\hspace{-2mm}
}

\foreach \x in {A, ..., Z}{%
	\expandafter\xdef\csname orcid\x\endcsname{\noexpand\href{https://orcid.org/\csname orcidauthor\x\endcsname}{\noexpand\orcidicon}}
}

\begin{document}

\title{Investigation of Data Deletion Vulnerabilities in NAND Flash Memory Based Storage\\
}
\author{Abhilash Garg\orcidA{}$^{1}$, Supriya Chakraborty\orcidB{}$^{1}$, Manoj Malik$^{2}$, Devesh Kumar$^{2}$,
Satyajeet Singh$^{2}$ and Manan Suri\orcidC$^{1}$ 
\\1. Department of Electrical Engineering, Indian Institute of Technology Delhi, India.
\\2. Defence Research and Development Organization, India.
\\manansuri@ee.iitd.ac.in
}

\maketitle

\begin{abstract}
Semiconductor NAND Flash based memory technology dominates the electronic Non-Volatile storage media market. Though NAND Flash offers superior performance and reliability over conventional magnetic HDDs, yet it suffers from certain data-security vulnerabilities. Such vulnerabilities can expose sensitive information stored on the media to security risks. It is thus necessary to study in detail the fundamental reasons behind data-security vulnerabilities of NAND Flash for use in critical applications. In this paper, the problem of unreliable data-deletion/sanitization in commercial NAND Flash media is investigated along with the fundamental reasons leading to such vulnerabilities. Exhaustive software based data recovery experiments (multiple iterations) has been carried out on commercial NAND Flash storage media (8 GB and 16 GB) for different types of filesystems (NTFS and FAT) and OS specific delete/Erase instructions. 100 $\%$ data recovery is obtained for windows and linux based delete/Erase commands. Inverse effect of performance enhancement techniques like wear levelling, bad block management etc. is also observed with the help of software based recovery experiments.
\end{abstract}


\section{Introduction}\label{intro}

In recent years, Flash memory has gained popularity in storage market with its increasing use in both embedded and standalone memory products. Embedded products include microcontrollers, SoCs while standalone products include solid state drives (SSDs), USB drives, Compact Flash (CF) cards, SD cards etc. NOR Flash is widely used in code storage applications due to its faster speed and random access capabilities whereas NAND Flash has its dominance in mass storage applications due to its high density/low cost per bit advantage \cite{micheloni2010nand}. Evolution of 3D NAND Flash memories has greatly improved the density and cost per-bit and enabled it as an undisputed NVM contender in the present day semiconductor memory market \cite{grupp2012bleak}. Over the last few decades, focus of the Flash industry has revolved around building storage media that are: (i) extremely dense (low-cost per bit), (ii) faster, (iii) low-power, (iv) have high cycling endurance, and (v) strong data retention capabilities \cite{lu2009future}. These properties have added value to the Flash memory products but at the same time they have led to vulnerabilities associated with data deletion and security \cite{jeong2007vulnerability, kim2013vulnerability, wei2011reliably}. Such data deletion vulnerabilities create serious concern for sensitive data (example- defense and strategic applications). 
Key contributions of this paper are:
\begin{itemize}
\item Identification and systematic categorization of different factors which result in data deletion vulnerabilities in Flash based storage. 
\item Supporting the assertions, by exhaustive software-recovery tool based data recovery experiments performed across multiple stored file types (texts, Images, Audio, Video, compressed, executables etc.) and filesystems (NTFS and FAT). 
\end{itemize}

This paper is organized as follows. Section \ref{section2} discusses the security relevant basics and system level organization of Flash memory technology. Section \ref{dv}, discusses multiple different reasons for data deletion vulnerabilities in Flash memory. Section \ref{dr}, discusses the approach of software based data recovery for different file-systems. Section \ref{drexperiments}, presents multiple experimental results of successful data recovery from USB NAND Flash storage media post-deletion. Section \ref{conclusions} presents the conclusion of the study. 
\begin{figure}
\includegraphics[scale=0.34]{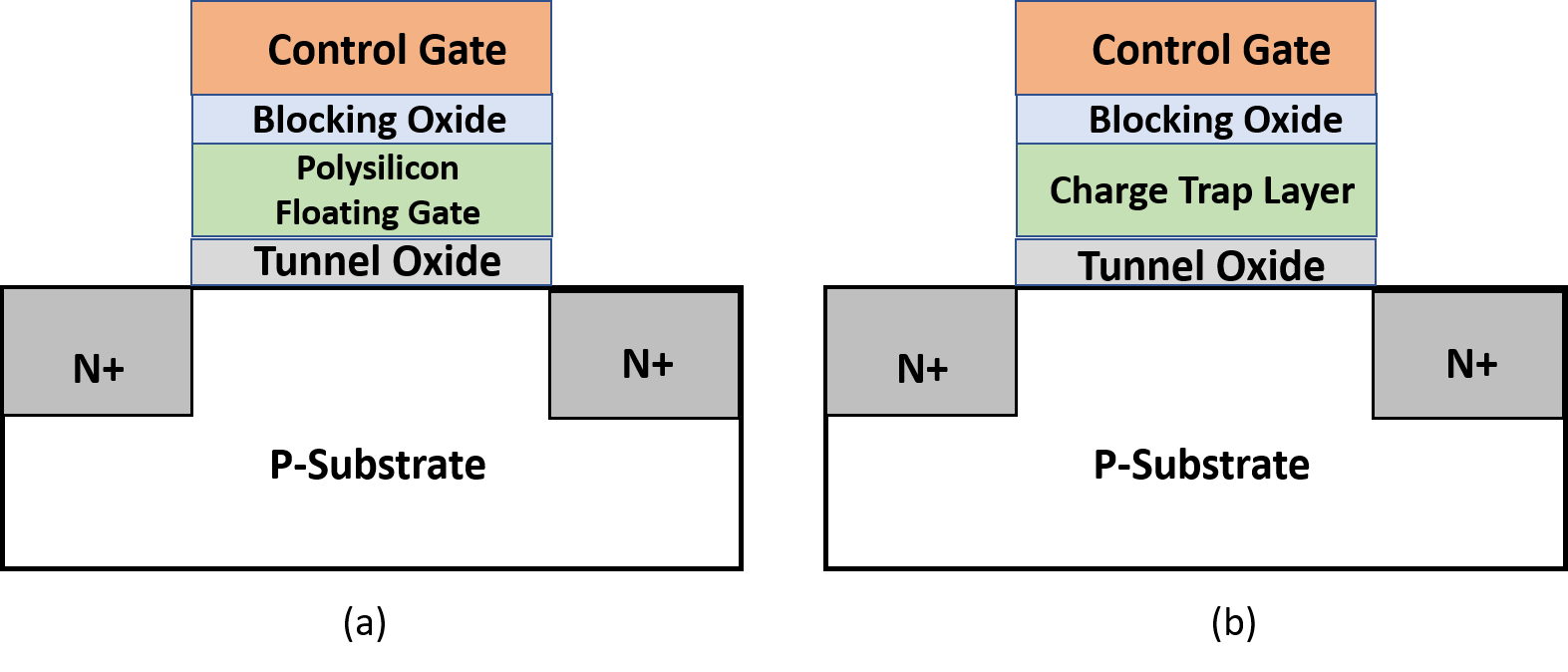}
\caption{\label{fig:FG_ONO} (a) Floating gate (FG) type Flash memory cell. (b) Charge Trap type Flash memory cell.} 
\vspace{-5mm}
\end{figure}

\section{Flash Memory Technology}\label{section2}
\subsection{Flash Memory Cell Basics}
Flash cell transistor, also known as the Floating-Gate (FG) transistor similar to conventional NMOS transistor device with a "Stacked Gate" configuration is shown in Fig.~\ref{fig:FG_ONO}(a). Top most layer in the FG stack structure is called as Control Gate (CG), it acts like a normal gate which can control the current passing through the transistor. Below the CG lies a blocking dielectric layer (BO), usually made up of layered oxide-nitride-oxide (ONO). The term blocking oxide or blocking layer is used because the function of this ONO layer is to prevent any charge carrier movement either from the CG or towards the CG from the layers below the CG. Below the BO lies a second gate, called as floating gate (FG), usually made up of poly-silicon \cite{flash_introduction}. The term floating-gate is used to denote this layer as it doesn't have any terminal or external contact to directly control its potential or apply any signal to it. Below the FG lies the thin tunnel oxide layer, usually made up of high quality silicon dioxide \cite{flash_introduction}. Additionally, in some type of emerging Flash memory, nitride (ONO) itself is used as charge trapping layer by performing engineering on the stack. This is called Charge Trap type Flash memory as shown in Fig. \ref{fig:FG_ONO}(b).
\begin{figure}
\centering
\includegraphics[width=0.45\textwidth]{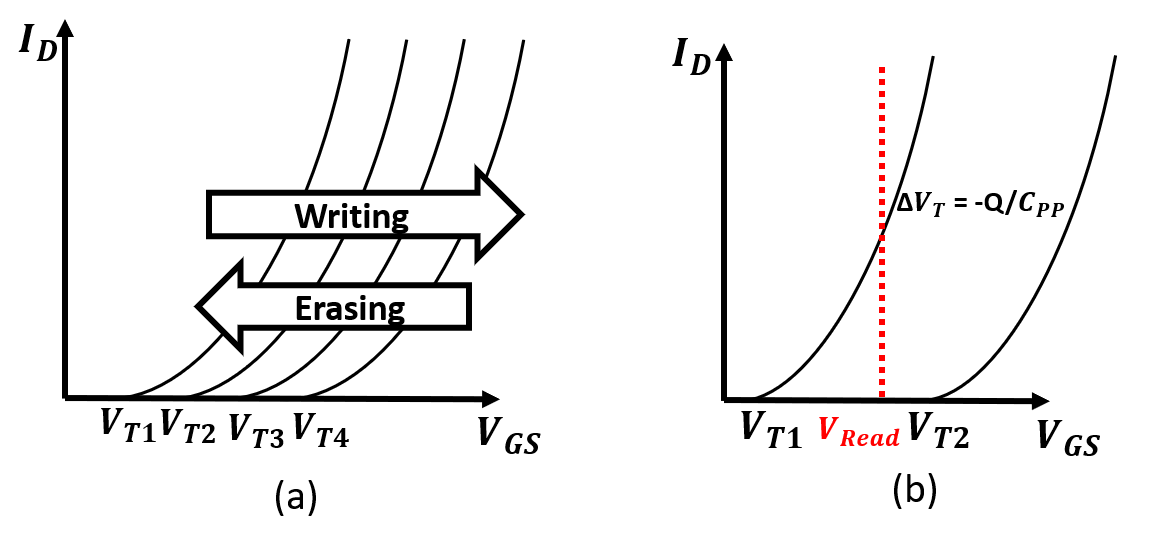}
\caption{\label{fig:FG_VT}(a) Threshold voltage modulation in FG Flash device during programming, (b) Read/Sense Operation in FG device. }
\vspace{-5mm}
\end{figure}
\subsection{Cell Programming Mechanism: Write, Erase, Read}
Basic underlying principle of FG Flash memory cell depends on sensing the relative change in threshold voltage of the device when charge is either intentionally trapped or de-trapped inside the charge-trapping layer. The process of modulating the threshold voltage of Flash cell by either injecting charges (\textit{Write}) or removing charges (\textit{Erase}) from the Flash charge trapping layers is defined as programming of the Flash memory cell (see Fig.~\ref{fig:FG_VT}(a)). Programmed state of a Flash cell can be validated by measuring the amount of current passing through the drain terminal of the cell on application of a fixed Control-gate voltage (Read Voltage). This process is defined as "reading". For a given read-voltage, negligible or $\sim$ zero drain current indicates that the cell threshold voltage is higher than applied read-voltage, and this state can be defined as the state storing logic '0'. If measurable current ($\sim$ \si{uA} to $\sim$ \si{mA}) flows from the drain terminal on application of a read-voltage it means an inversion channel has been formed. This turn-on state can be defined as the state storing logic '1'. Read operation based on threshold voltage sensing is illustrated in Fig.~\ref{fig:FG_VT}(b).
\\Different physical mechanisms and techniques exist in literature for charge injection into the FG \cite{pavan1997flash}. Two most common include: (i) Channel Hot Electron Injection (CHE), and (ii) Fowler-Nordheim (FN) tunneling. CHE mechanism relies on first accelerating the channel electrons using a lateral electric field (between source and drain) and then force injecting the energetic charge carriers to the floating gate, through the tunnel oxide, with the help of vertical electric field (between channel and CG) (see Fig. ~\ref{fig:che,fn}(a)). 
\\In Fowler-Nordheim (FN) tunneling, a sufficiently high vertical electric field between the CG and channel region is maintained  (Fig. ~\ref{fig:che,fn}(b)). This leads to steep energy band diagram for the oxide region; thus significantly increasing the probability of electron tunneling through the tunnel-oxide energy barrier. NOR Flash, mainly uses CHE injection mechanism for Write operation, and FN tunneling for Erase operation \cite{flash_introduction}. While NAND Flash uses FN tunneling for both Write and Erase \cite{micheloni2010nand}.

\begin{figure}
\includegraphics[width=0.55\textwidth]{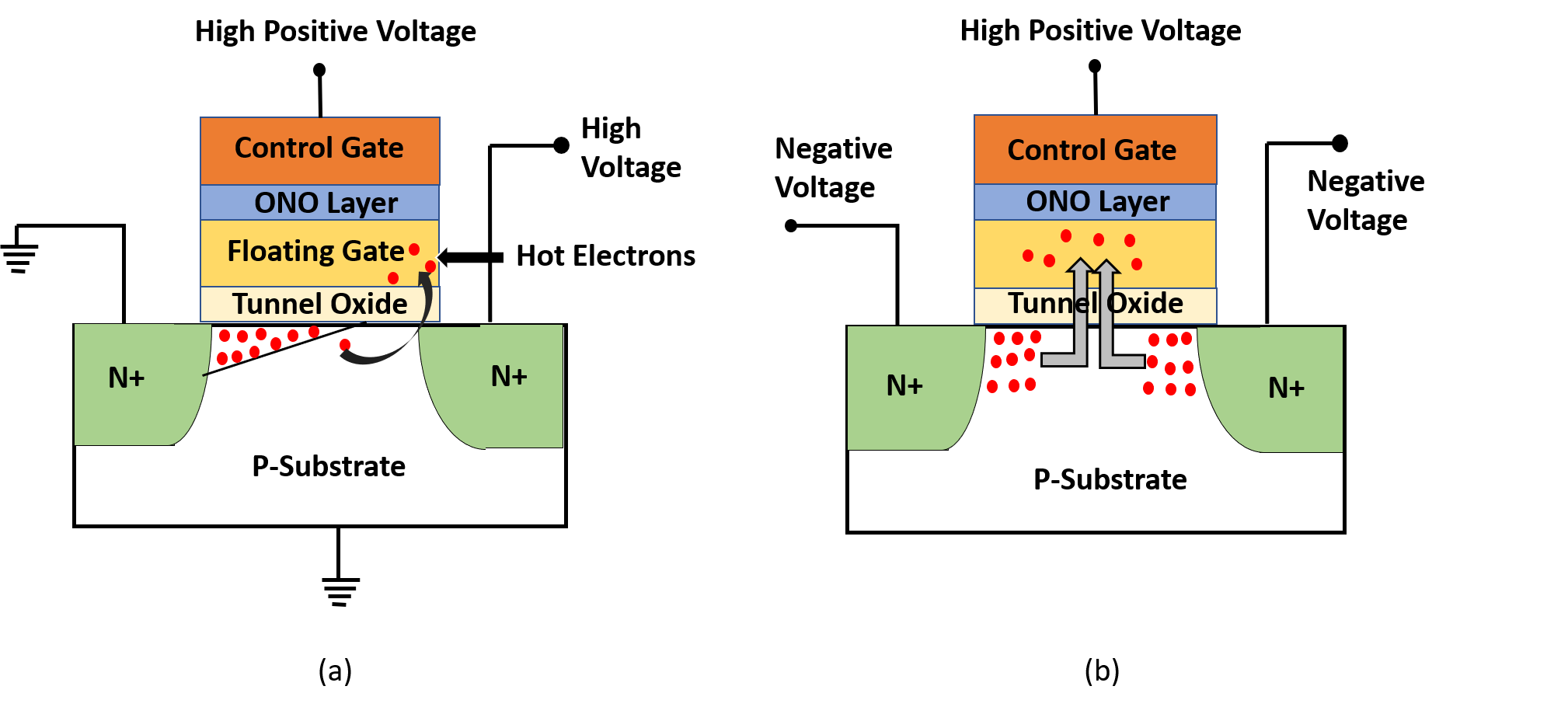}
\vspace{-6mm}
\caption{\label{fig:che_fn}Programming mechanisms for FG Flash devices: (a) CHE/HEI, and (b) FN tunneling.} 
\label{fig:che,fn}
\end{figure}
\begin{figure}
\centering
\includegraphics[scale=0.5]{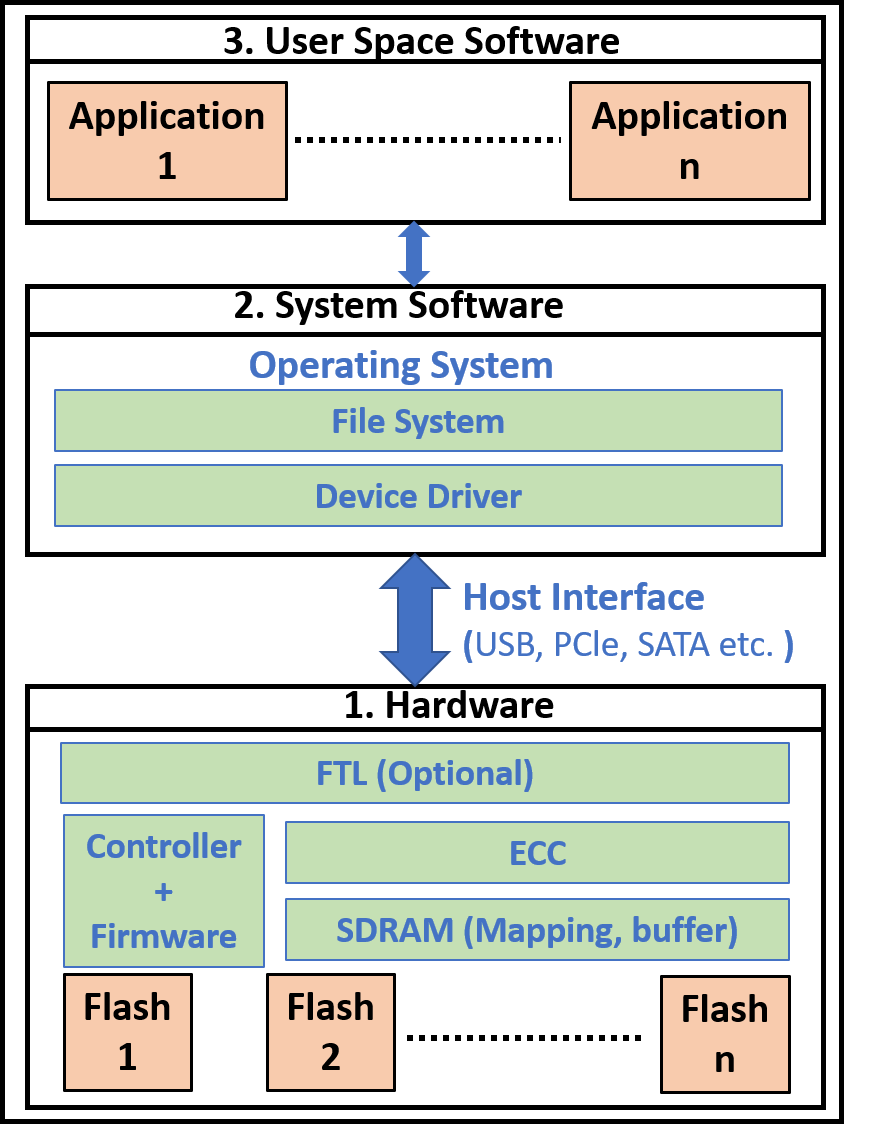}
\caption{\label{fig:arch} System organization and abstraction layers in Flash memory based storage media.}
\vspace{-5mm}
\end{figure}

\subsection{System Level Organization}{\label{system organization}}

System organization for a Flash memory based storage media can be divided into following three major components or layers of abstraction (Fig.~\ref{fig:arch})  \cite{reardon2013sok}: 

\begin{enumerate}
    \item Hardware (Storage medium, Controller, Host Interface)
    \item System Software (Device drivers, Firmware, File System)
    \item User-space software (i.e. applications like File Managers/File Browsers)
\end{enumerate}
Every abstraction layer between end-user and the memory device array/chip makes the Flash structure more complex. Complexity in structure results in data-deletion/sanitization related vulnerabilities which are discussed in detail in the following section (section \ref{dv}).  

\section{Data Deletion Vulnerabilities}
\begin{figure}
\includegraphics[scale=0.25]{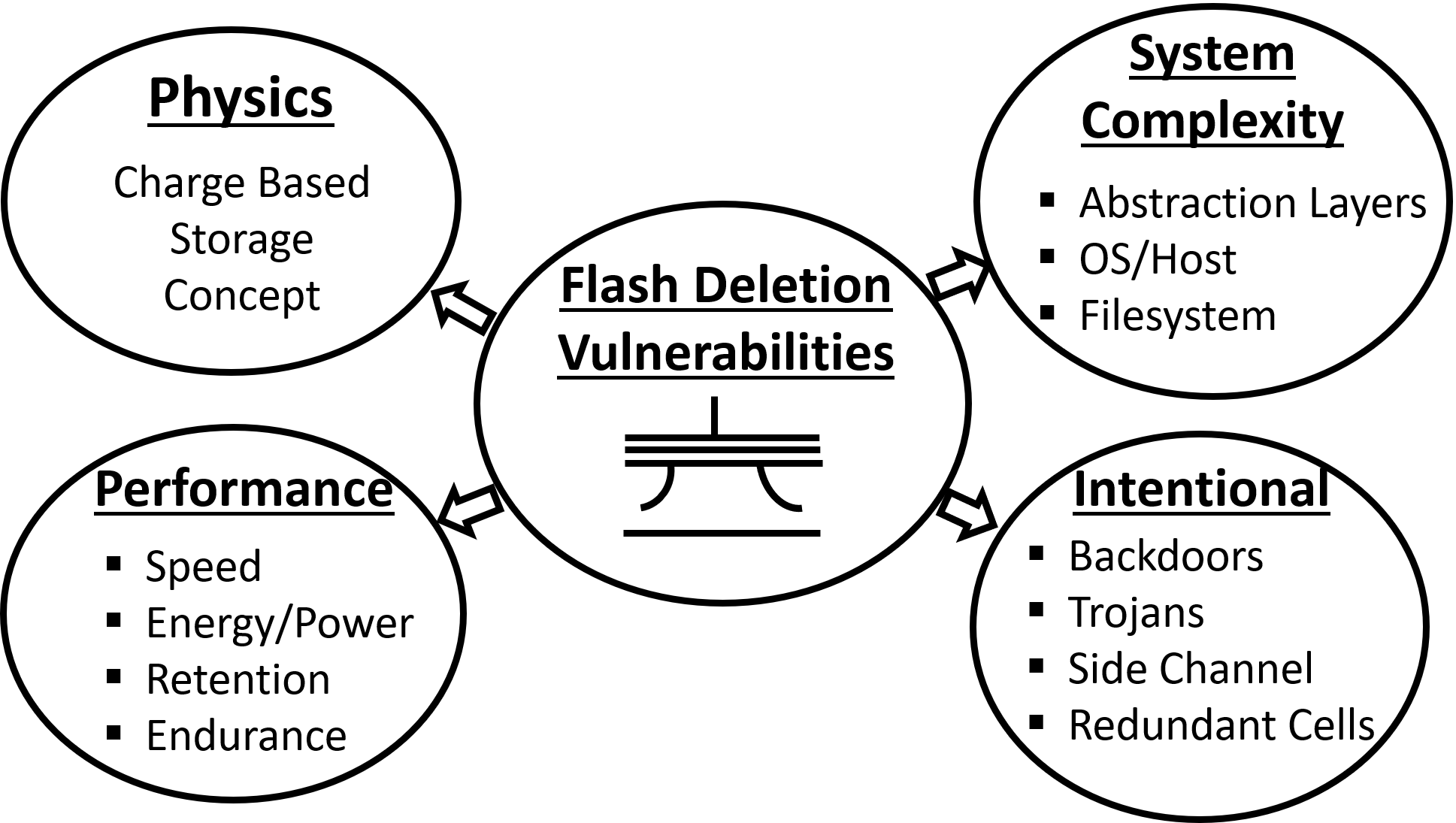}
\caption{\label{fig:deletion_problems}Categorization of Flash data-deletion/sanitization vulnerabilities in different categories.}
\vspace{-3mm}
\end{figure}

{\label{dv}}
In this section, we analyze a comprehensive list of different possible data-deletion/sanitization vulnerabilities in Flash memory based storage media, across all abstraction layers.
We group all vulnerabilities in four different categories (Physics, Performance, System Complexity, and Intentional, Fig.~\ref{fig:deletion_problems}) based on the underlying technical motivations and reasons.
\vspace{-5mm}
\subsection{Physics Based Reasons}{\label{physics based}}

\begin{figure}
\centering
\includegraphics[scale=0.32]{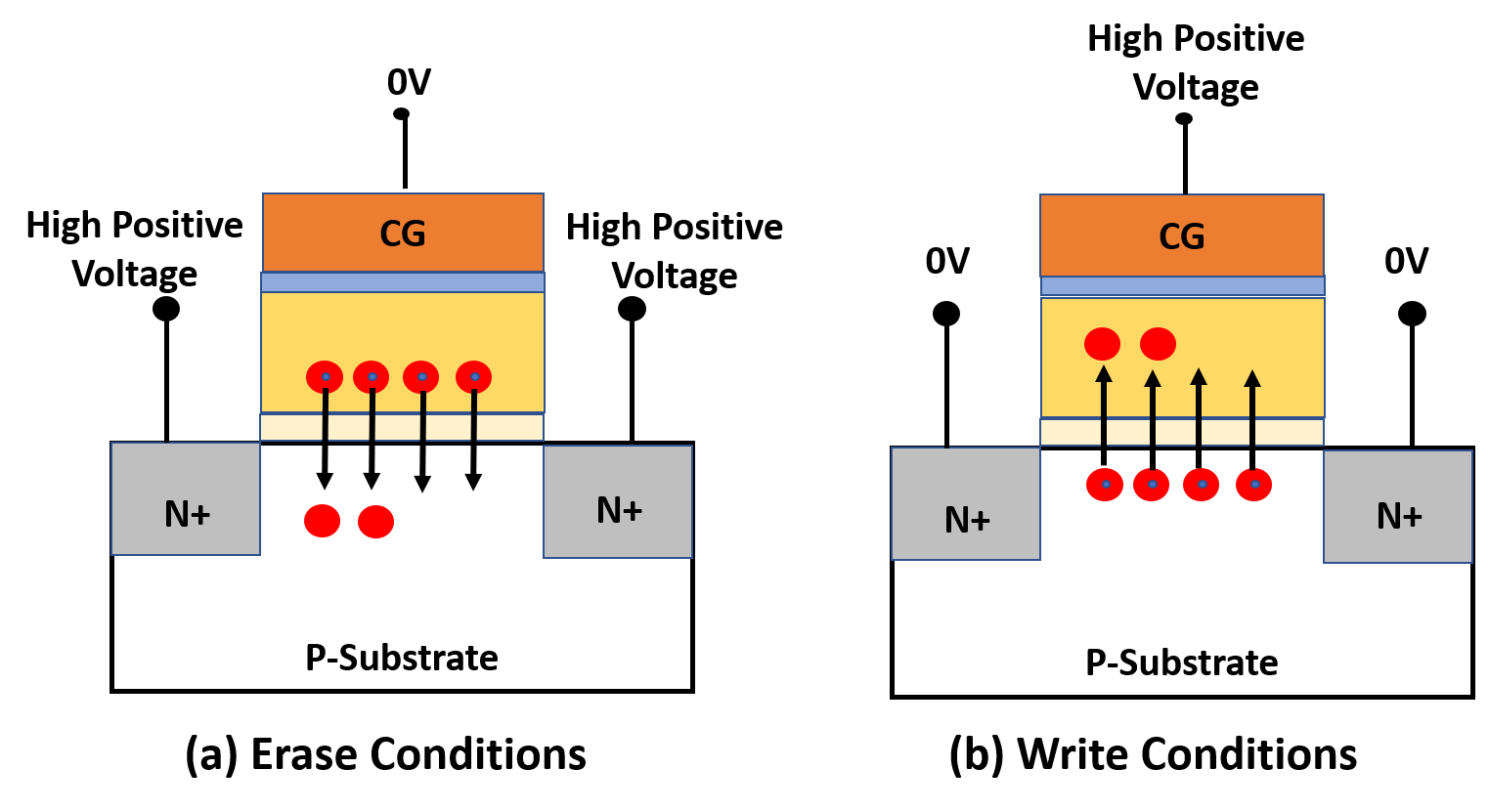}
\caption{\label{fig:fg_charges}Write/Erase operation state definition based on charge trapping/de-trapping in FG Flash memory device.}
\vspace{-5mm}
\end{figure}

Major reason giving rise to deletion vulnerability in Flash storage is its underlying charge based information storage principle. Stored state of a FG Flash device depends upon the relative difference in its threshold voltage which is proportional to the amount of charge stored in the charge trapping layer (Fig.~\ref{fig:fg_charges}). It is worth noting that charge carriers (electrons/holes) will behave quantum mechanically and in terms of populations. In other words, when FG Flash device is programmed to state '1' a certain finite number (population) of electrons are injected in the charge trapping layer volume (and not just a single electron). When the device undergoes Erase operation the threshold voltage based sensing/reading mechanism may indicate successful Erase or switching to state '0' even when the entire population of electrons that was injected is not completely de-trapped and few electrons are still left behind. The quanta of charge carriers trapped or de-trapped from the charge-trapping layer will depend upon the device dimensions and the exact programming conditions (voltages, duration, resultant E-fields) applied during Write/Erase operations. Erasing or de-trapping of charge back to substrate will happen similar to a capacitive decay leaving behind some remanent charges \cite{rao2005cryptographic, skorobogatov2005data}. There is no guarantee that the OS/data-sheet defined digital Erase instruction ensures zero trapped charges in the charge-trapping layer post completion of the Erase process. The user has no control over the analog parameters (voltages/duration) of the programming conditions applied. In case of modern Multi-level-cell (MLC) Flash memories, this problem becomes even more pronounced and need to be addressed \cite{lin2018achieving}. 
These factors may lead to a so called charge remanence post-Erase. Remanent charges or charge contrast in Flash cells can be observed with advanced nanomaterial characterization techniques like: Scanning Electron Microscopy (SEM) and Scanning Probe Microscopy (SPM) (AFM, SCM) \cite{front_back_approach, courbon2016direct, sample_preparation}. Based on the amount of residual charges, detailed physics based/empirical models can be built, which can predict the previous state of the Flash device prior to the launch of an Erase or Overwrite instruction. Remanent charges allow to distinguish between previously programmed and non-programmed FG devices in terms of threshold voltage, even after one hundred Erase cycles \cite{rao2005cryptographic}. Thus adversaries may exploit such residual charge sensing or charge-contrast sensing to recover and decode deleted/Erased/Overwritten data.

\subsection{Performance Based Reasons}
\begin{figure}
\includegraphics[width=0.47\textwidth]{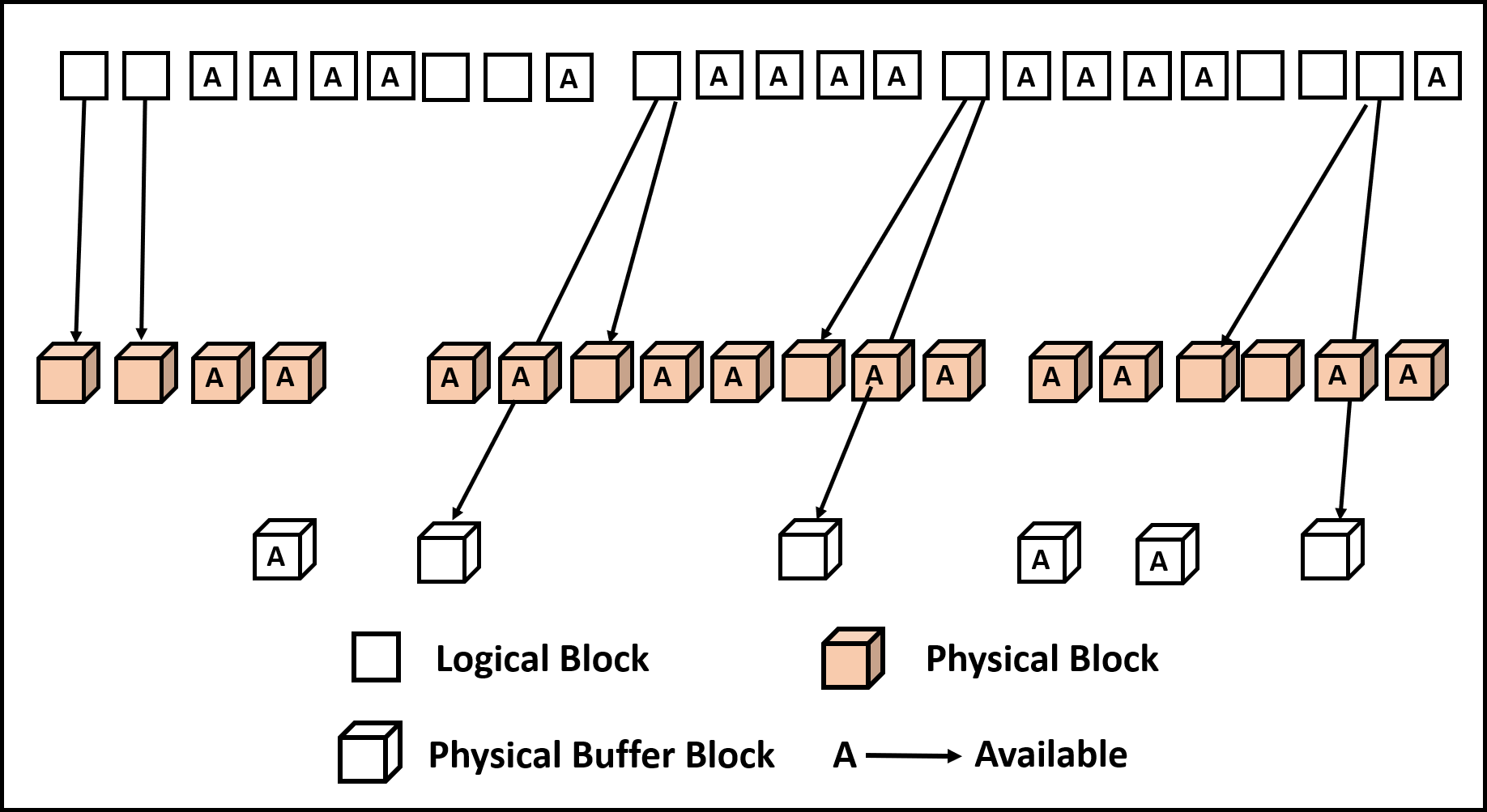}
\caption{\label{fig:FTL} Mapping of logical blocks to physical blocks inside FTL \cite{micheloni2010nand}.}
\end{figure}

\begin{figure}
\includegraphics[width=0.47\textwidth]{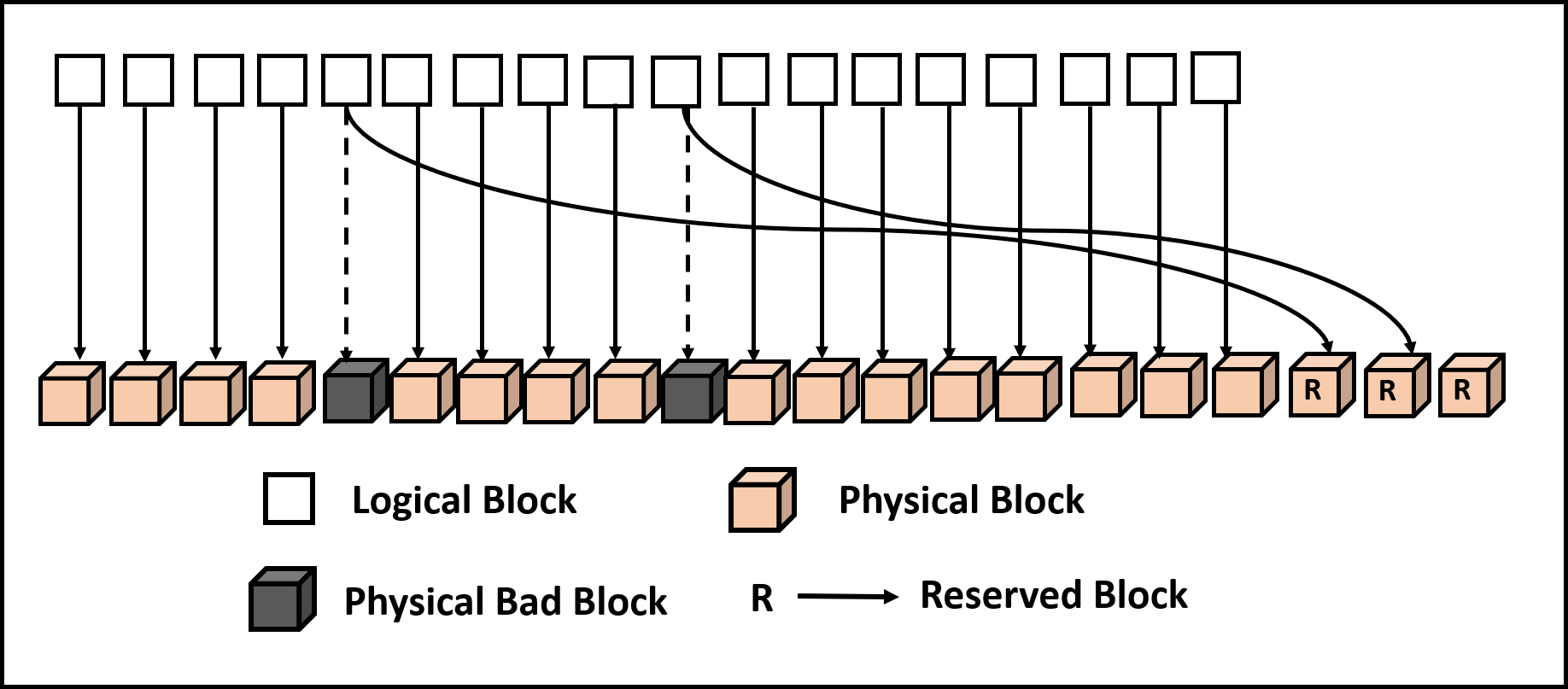}
\caption{\label{fig:bb} Replacement of bad physical blocks with fresh reserved blocks \cite{micheloni2010nand}.}
\vspace{-5mm}
\end{figure}

{\label{performance based}}
Second primary category of reasons for deletion vulnerabilities can be attributed to Flash storage media's performance enhancement related factors. Performance optimization tricks may lead to data security trade-offs in some cases. Key performance metrics of interest include (Fig.~\ref{fig:deletion_problems}): (i) programming latency, (ii) programming energy/power, (iii) data retention, (iv) endurance and (v) error management. 
\\\textbf{Programming latency and energy/power:} In commercial NAND Flash based storage media, in order to optimize the Flash Write/Erase performance (i.e. speed, power), the cell threshold sensing margin is defined such that even partial removal or addition of charge can be sensed as successful completion of Erase/Write instructions. These factors may lead to a so called charge remanence post-Erase. As described in section \ref{physics based}, this remanent charge can be sensed in the form of charge contrast by using nanomaterial characterization techniques to predict previously stored state of the Flash device. 
\\\textbf{Retention}: High retention in Flash devices ensures that a residual trapped charge post Erase is very less likely to be lost and thus enhances the probability of information recovery even after long time.
\\\textbf{Endurance}: Flash memory devices often suffer from low endurance (less than 10,000 Write/Erase cycles \cite{pavan1997flash}). To virtually enhance the endurance of Flash memory based storage media, File translation Layer (FTL) provide techniques like wear-levelling, garbage collection, bad block management etc. which are implemented inside memory controller \cite{ftl}. These techniques use address mapping to map logical blocks to physical buffer blocks of memory with the help of file translation tables as shown in Fig.~\ref{fig:FTL} \cite{ftl_2002}. Whenever host issues instruction to write data on a specific address, the memory controller may dynamically map same logical address to a different physical address, in order to uniformly distribute cycling. Such mapping information is stored in a translation table. Thus wear levelling helps in intelligent distribution of Write/Erase cycles across different Flash memory devices on the chip to improve endurance. As a consequence of wear levelling, previously written data may still exist in physical Flash devices, giving rise to data remanence problem. Whenever a physical block reaches its endurance limit then its mapping gets replaced with the fresh reserved physical block (Fig.~\ref{fig:bb}) with the help of bad block management module, but data in old physical blocks remain intact and can be accessed/retrieved leading to data-deletion/sanitization vulnerability. 

\subsection{System Complexity Based Reasons}{\label{system complexity}}
Major factors causing system complexity related data-deletion/sanitization vulnerabilities in Flash memory based storage media include (Fig.~\ref{fig:deletion_problems}): (i) design of OS/Host, and (ii) Filesystem. 
As described in section \ref{system organization}, NAND Flash memory based storage media have multiple abstraction layers/interfaces above raw Flash devices which makes storage system complex (Fig.~\ref{fig:arch}). These layers include: (i) hardware, (ii) system software, and (iii) user space software. These abstraction layers/interfaces make storage media suitable for end-user application and Host interaction with the Flash devices but it adds more stages of data-deletion/sanitization vulnerabilities. Vulnerabilities caused by the hardware abstraction layer are discussed in section \ref{physics based} and section \ref{performance based}. System software related data-deletion/sanitization vulnerabilities arises from OS/Host and Filesystem. 
\\\textbf{OS/Host}: Design of OS or utility programs to delete data are not upto the mark (see section \ref{drexperiments}). OS can provide local buffer/file retention facility, instead of deleting files it can move them to a holding area (example Recycle Bin) from where files can easily be recovered back. OS generally removes entries from filesystem directory, upon execution of file deletion command, by only `marking' the previously occupied space as `unused' and `free' for further writing. Actually data may still remain in the physical Flash devices for long duration  \cite{castiglione2011automatic}.
\\\textbf{Filesystem}: It generally stores metadata (e.g. path names, last modified time, etc.) of deleted files. By using this metadata files can be recovered back using easily available software recovery tools \cite{MFT}. File recovery using filesystem has been described in section \ref{dr}. Reformatting an entire storage media or a partition implies the destruction of filesystem metadata, but does not guarantee that the data present in the formatted area is completely erased at the physical Flash device charge level (deletion vulnerabilities due to hardware abstraction layer may still cause data remanence) \cite{lee2008secure}. 

\subsection{Intentional Reasons}{\label{intentional}}
Vulnerabilities discussed in sections \ref{physics based}-\ref{system complexity} are not directly implemented procedures for information leakage. They are either consequences of some physical phenomenon or consequences of some performance enhancement technique implemented inside NAND Flash based storage technology. However, an adversary may intentionally introduce vulnerabilities without user's knowledge. Major intentional vulnerabilities include: (i) backdoors, (ii) trojans, (iii) side channel attacks, and (iv) provision of redundant memory cells.
\\\textbf{Backdoors and Trojans}: These are intentionally introduced threats in semiconductor ICs to provide unauthorized access to control or information of a system. They may change the functionality of memory circuits when activated. Trojans can be activated internally (by inbuilt circuit logic) or externally (by using antennas, sensors etc.). Such vulnerabilities can only be introduced during design/fabrication phase of the IC or the PCB \cite{trojan}. Detection and protection against such provisions are very unlikely for a normal user. Backdoors have been successfully discovered even in military chips in the past \cite{backdoor}. 
\\\textbf{Side Channel Attacks}: Information from Flash based storage media can be recovered/retrieved successfully by exploiting external or indirect parameters such as timing information, optical, acoustic, power etc. Information can also be extracted from Flash based storage media by using bumping attacks \cite{skorobogatov2010flash}. Apart from unauthorized data extraction, side channel attacks may also be used to modify, manipulate or destroy the stored data. For instance, content of Flash devices can be altered successfully by locally heating them \cite{skorobogatov2009local}. These factors create serious threat when confidential data is stored inside Flash storage media (example: cryptographic keys). 
\\\textbf{Redundant cells}: Recently, as Flash memory density has drastically increased, it has led to significant on-chip silicon area savings for Flash arrays of fixed sizes. The saved die-area can be utilized by adversaries to introduce malicious circuits, IPs or even redundant Flash memory cells. Since silicon area remains the same, the malicious circuit/device addition bears no extra cost or area on the Flash memory chip. 
Redundant cells may contain copy of user data and can only be accessed by using special adversary instructions. An adversary having access to this special instruction set to access the storage media can easily hack user data without their knowledge.

Flash based storage technology follow increasingly standardized and cross platform compatible protocols and interfaces (ONFI, SCSI, SATA etc.). If an adversary manages to hack one type of Flash storage media then replicating the methodology across different commercial storage media's would be probable.
\section{Data Recovery in Flash Memory}{\label{dr}}

\begin{figure}
\centering
\includegraphics[width=0.45\textwidth]{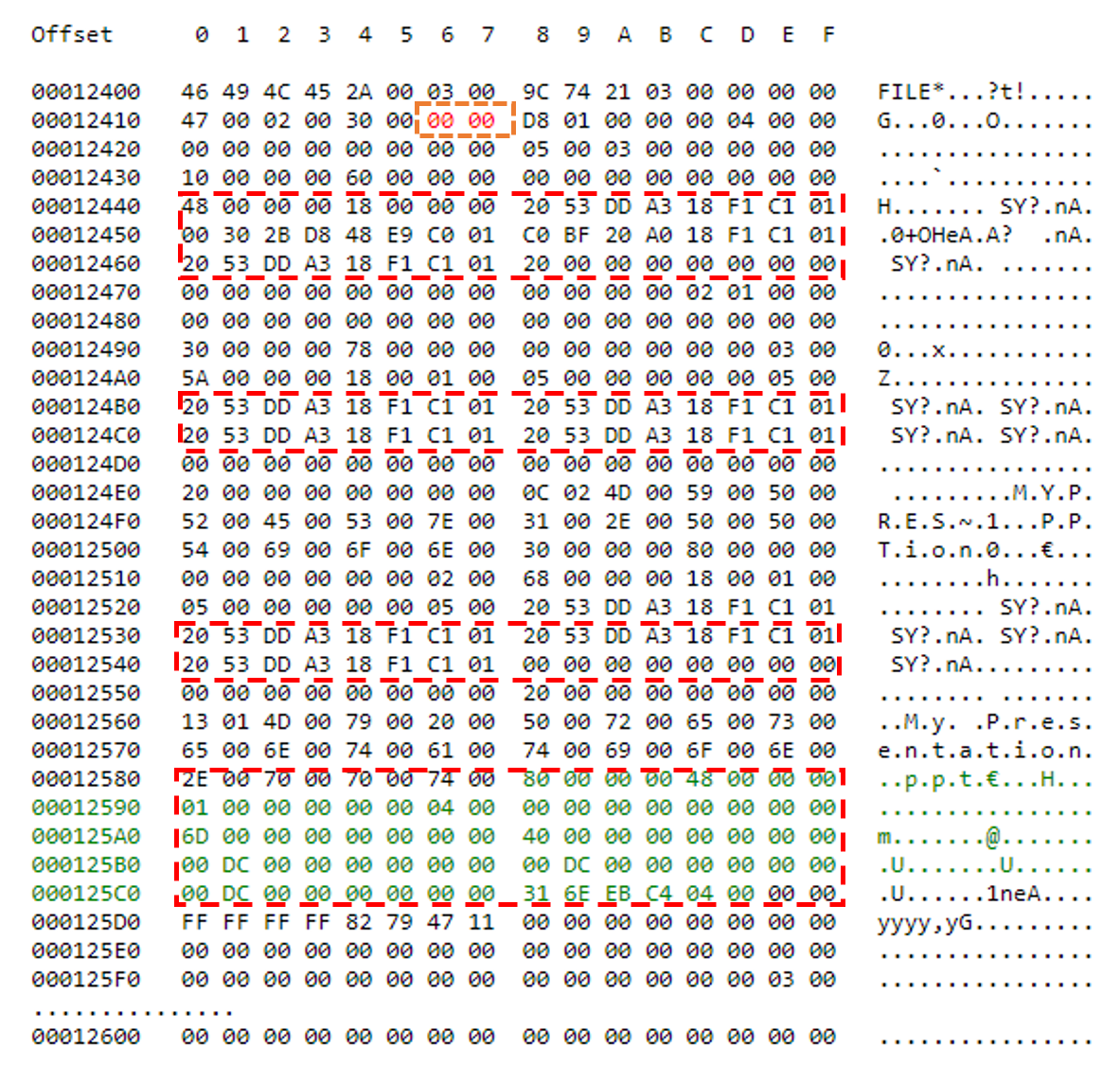}
\caption{\label{fig:MFT} Example of MFT structure on NTFS filesystem which helps in data recovery \cite{MFT}.}
\vspace{-5mm}
\end{figure}

Data can be successfully recovered from Flash based storage media by exploiting vulnerabilities described in section \ref{dv}, in following ways: (i) with the help of software recovery tools \cite{MFT}, (ii) by building comprehensive embedded test environment \cite{wei2011reliably}, and (iii) by direct charge/charge contrast measurement of Flash devices \cite{sample_preparation}.
\\Data recovery softwares exploit vulnerability of filesystem for successful recovery post-deletion. Different data-recovery tools are available which are helpful in recovering deleted/corrupted files from a normal/damaged (Flash and HDD) storage media. Where the main intent behind building these tools was to help users recover their valuable files if accidentally lost, they can always be misused for recovering data in unauthorized manner that has been deleted by an end-user.
\\File Recovery softwares work on a systematic process which involves scanning of deleted entries in the disk space, cluster chain definition for deleted entries and finally recovering data using these cluster chains. These deleted entries are stored in Root Folders on FAT12, FAT16, FAT32 or in Master File Table (MFT) on NTFS, NTFS5 filesystems. Different filesystems have different structure for files/folders but basic attributes (name, size, creation and modification time/date, existing/deleted status, etc.) are always present in all of these filesystems. Record of deleted entries are stored differently for different filesystems. In NTFS, file header provide special attribute for deleted files/folders and in FAT ASCII symbol 0xE5 (229) is used to mark files/folders deleted. By scanning MFT record for these attributes information about deleted files/folders is extracted. First 42 Bytes (starting from 0x00) in MFT contains file record header in which Flag field (2 Bytes) provides information about whether the file is deleted or still in-use. If the flag bit is set to `1' then file is in-use otherwise deleted. MFT also contains file information attributes (creation and modification time/date, size, name of file) starting from address 0x48 (68 Bytes) as shown in Fig.~\ref{fig:MFT}. Non Resident Data attribute (starting from 0x188) in MFT helps in finding compression unit size, allocated and real size of attribute and data runs. Many more information fields about file/folder exist in MFT which we have not discussed here, this can be found in \cite{MFT}.

In this way all the information about the deleted files can be extracted from the boot record or MFT of filesystem record.
Next step is to define cluster chains for deleted files. Complete storage media is scanned to define cluster chain. Data Runs give information about the location of the file clusters in NTFS filesystem. Data Runs are decrypted by using suitable methods as explained in \cite{MFT}. 
\\After obtaining cluster chains definition successfully, next step is to recover these cluster chains (read and save) for the successful recovery of deleted entries. Therefore, cluster offset is calculated using standard formula. Once the offset is known then deleted data is copied to the new file starting from the calculated offset to last address (file size minus number of copied clusters multiplied by cluster size).
\\One research provided 3 extreme cases where data was recovered even after destructing the USB Flash memory with over-voltage, soaking in water, incinerating in petrol, stomping and hammering \cite{phillips2008recovering}. Another work in literature have used FPGA based test environment and illustrated the formation of multiple copies of same data in Flash SSDs, resulting in data recovery after deletion \cite{wei2011reliably}. Forensic data recovery from Flash based storage media using different methods have also been shown by \cite{breeuwsma2007forensic}.

\section{Our Recovery Experiments and Results}{\label{drexperiments}}
 
To first hand investigate data remanence and data deletion vulnerabilities described in section \ref{dv}, we designed and performed extensive software data-recovery experiments. We used commercial NAND Flash USB media for the experiments. The experiments conclusively proved the failure of vendor and OS defined delete/Erase commands to reliably delete/sanitize target information of different formats.
\\New 16 GB (14.5 GB usable) and 8 GB capacity commercial NAND Flash USB storage media were used. For the sake of generality; (i) Windows 10 and Linux Ubuntu 14.04 OS were used, and (ii) USB storage media were initially formatted to both NTFS and FAT32 filesystems to perform the data recovery experiments. Recuva and EaseUS recovery software tools were used for recovering the deleted/Erased data.
\\As the USB storage media were new, initially, there is no user-data present on the media except the user firmware/application files. Firstly, user sample data of different file formats (text, images, audio, compressed, code etc.) and varying file sizes is written on to the USB media. Then these sample data files are deleted using different type of delete/Erase instructions available to the user. These instructions include OS specific 
deletion commands. For example, Windows offers: simple delete, Shift+Delete (permanent delete) and Format. Linux offers: simple delete (sudo rm -rf filename) and shred (sudo shred -n 1 -v /dev/sdb).
In order to improve the quality of the recovery results we performed multiple trials of recovery experiments with varying file sizes each time and varying sample data files written in the first place. Data recovery with the help of software recovery tools was attempted, post the application of different OS specific delete/Erase commands. 
Table ~\ref{table:recovery} provides a summary of all software tool based data recovery experiments. Contrary to user expectation and perception the data was easily recovered in several cases thus raising a cause for serious concern. 
Following sub-sections describe in detail the data-recovery results and observations.
\begin{table}[t!]
\renewcommand{\arraystretch}{1.5}
\caption{Software data recovery statistics for different data deletion instructions and file types.}
\begin{center}
\begin{tabular}{|c|c|c|c|c|c|}
\hline
 &  & \multicolumn{3}{c|}{\textbf{Percentage Data Recovery}}\\
\cline{3-5}
\textbf{S.No.} & \textbf{File Type} & \multicolumn{2}{c|}{\textbf{Windows 10 Home Basic}} & \textbf{Ubuntu}\\
\cline{3-5}
& & \textbf{Shift+delete} & \textbf{Format} & \textbf{Sudo rm -rf} \\
\hline
1 & Documents & 100 $\%$ & 100 $\%$ & 100 $\%$\\
& (Text Files) & & &\\
\hline
2 & Image Files & 100 $\%$ & 100 $\%$ & 100 $\%$\\
\hline
3 & Audio Files & 100 $\%$ & 100 $\%$ & 100 $\%$\\
\hline
4 & Video Files & 100 $\%$ & 100 $\%$ & 100 $\%$\\
\hline
5 & Compressed & 100 $\%$ & 100 $\%$ & 100 $\%$\\
\hline
6 & Executable  & 100 $\%$ & 100 $\%$ & 100 $\%$\\
& Files & & & \\
\hline
\end{tabular}
\label{table:recovery}
\end{center}
\vspace{-5mm}
\end{table}
\subsection{Data Recovery on Separate Storage Media}
\begin{figure}
\centering
\subfigure[]{\includegraphics[width=0.45\textwidth]{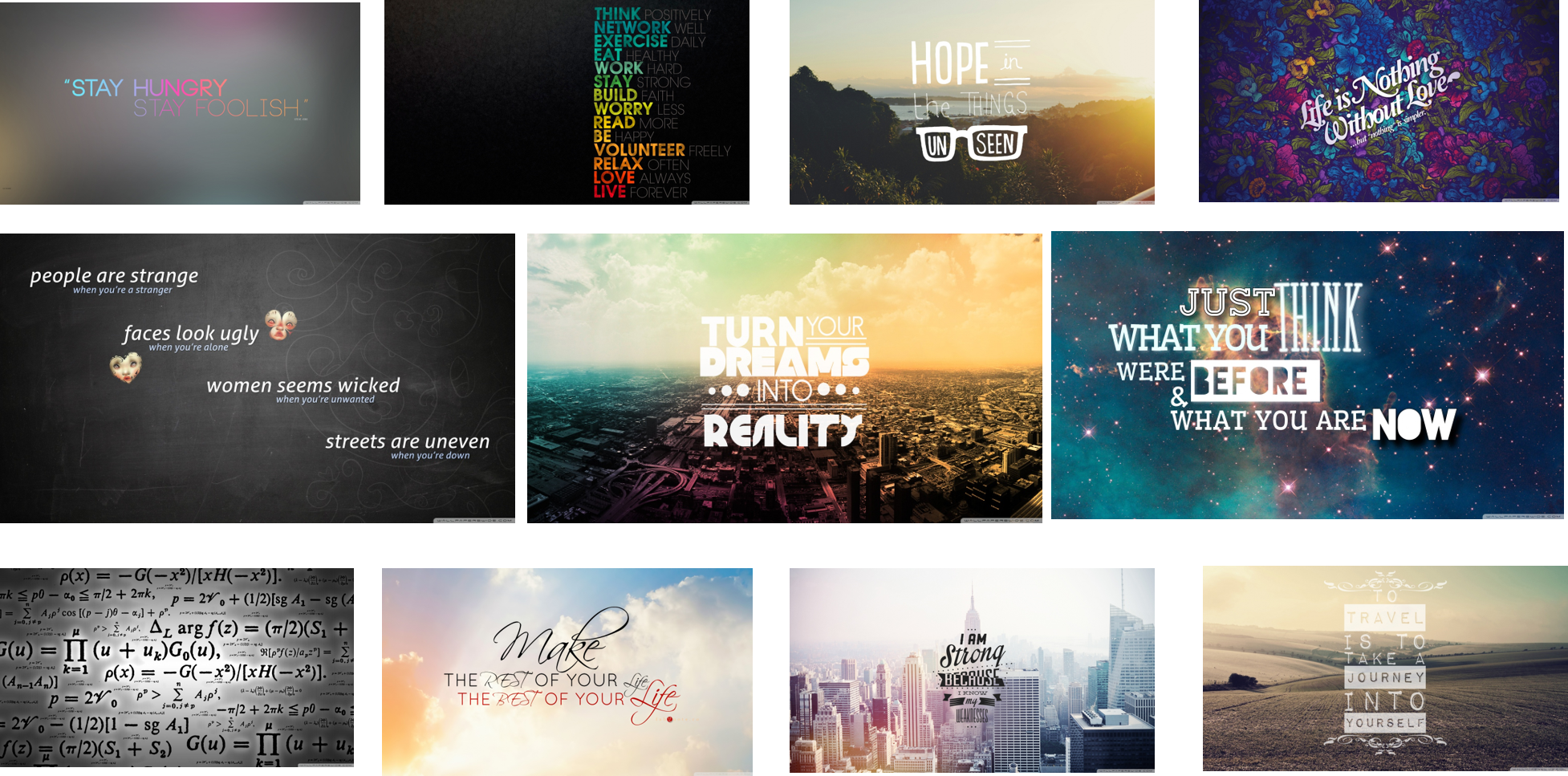}}\label{recovered_images}
\subfigure[]{\includegraphics[width=0.45\textwidth]{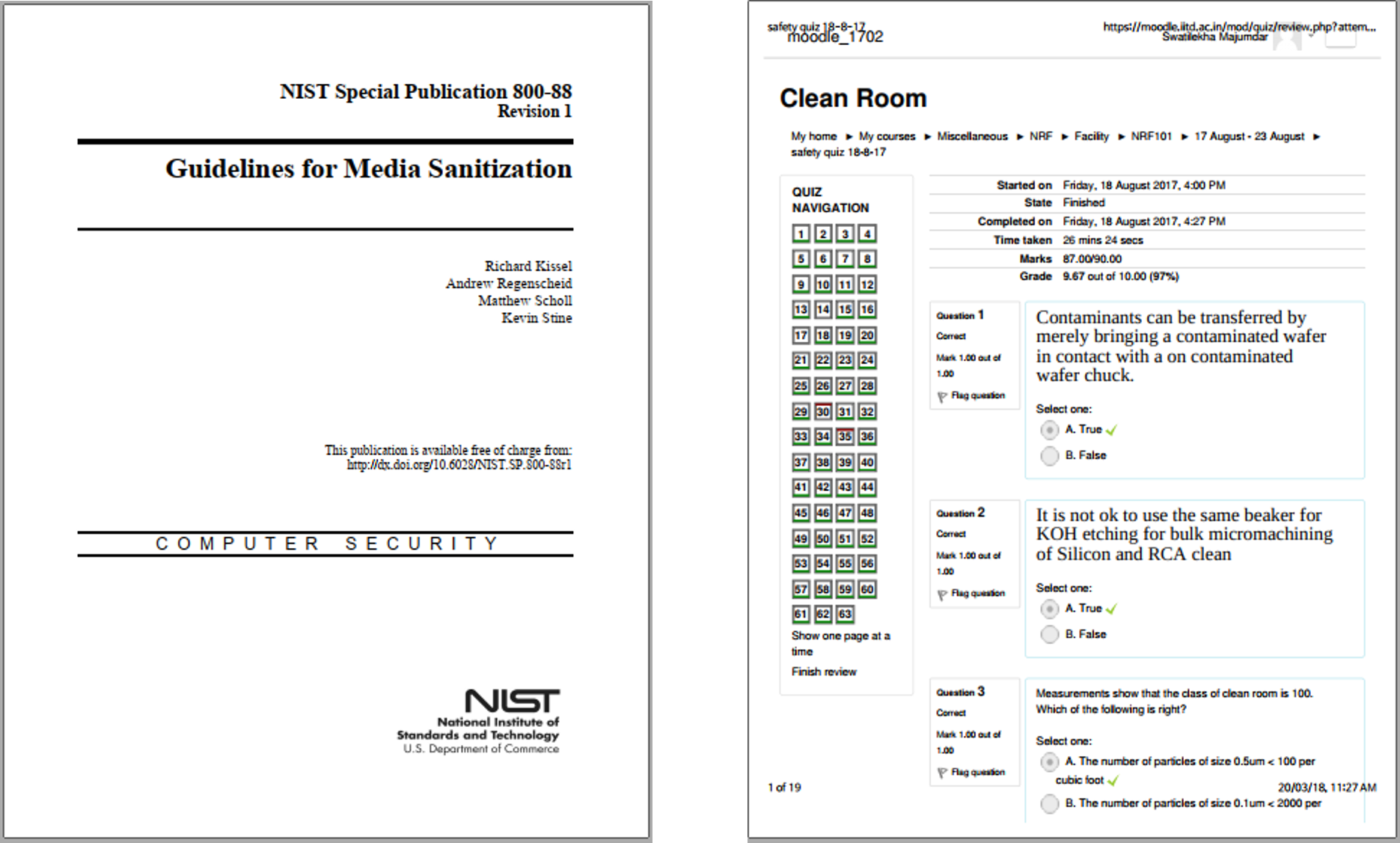}}\label{recovered_pdf}
\caption{Successfully recovered files after deletion (a) Recovered Image files. (b) Recovered text (pdf) files. All files are readable.}
\label{fig:recovered_images}
\vspace{-5mm}
\end{figure}

In these experiments the recovery tool was asked to dump the recovered data on the Host PC/Laptop's storage. 
In one of the experiments, a total 57.6 MB data was written to the 8 GB commercial NAND Flash USB storage media. Data was in the form of image files (.jpg, .png, .bmp, .tiff, etc.), documents (.doc, .docx, .ppt, .pdf, .xlsx, .rtf, . txt etc.), audio (.mp3) files, video files (.avi, .flv, .mp4, etc.), compressed files (.rar, .zip, .tar, etc.) and code files.  After writing into the USB storage media, these files were permanent deleted using Shift+Delete as claimed by the Windows OS. After which recovery software was used to recover the deleted files. All 57.6 MB of data was recovered successfully without any distortion in file format. Out of the recovered 57.6 MB, image files of 3.42 MB, and 1.08 MB of text (pdf) files, as recovered (post-deletion) are shown in Fig.~\ref{fig:recovered_images}.
\\After unsuccessful deletion of data by shift+delete command, quick format option in windows 10  was tried to permanently delete data from the same USB storage media. After deleting data by using quick format command, we used recovery software to check for the permanent deletion of data but this time deep scan was performed which was more time consuming. Again entire amount of data was successfully recovered without any distortion.
\\After successful recovery of data, full format option was tried to permanently delete data from same USB storage media. After running full format command, recovery software was run in deep scan mode to check for data remanence but they failed to recover any data. We suspect that this might be due to the limitations of data recovery softwares used because successful data recovery has been shown in literature by using customized embedded recovery setup \cite{phillips2008recovering}. 
\\We also tested Linux deletion commands to delete the data using normal delete (sudo rm -rf filename) and shred (sudo shred -n 1 -v /dev/sdb). Post normal-delete command, the data was successfully recovered by using recovery software.  However, no data was recovered after shred delete command. We suspect that data may be recovered successfully by making customized embedded recovery setup and by using invasive microscopy based techniques such as AFM/SCM/SEM etc \cite{courbon2016direct},\cite{sample_preparation}.

\subsection{Data Recovery on Same Storage Media}
\begin{figure}
\includegraphics[scale=0.28]{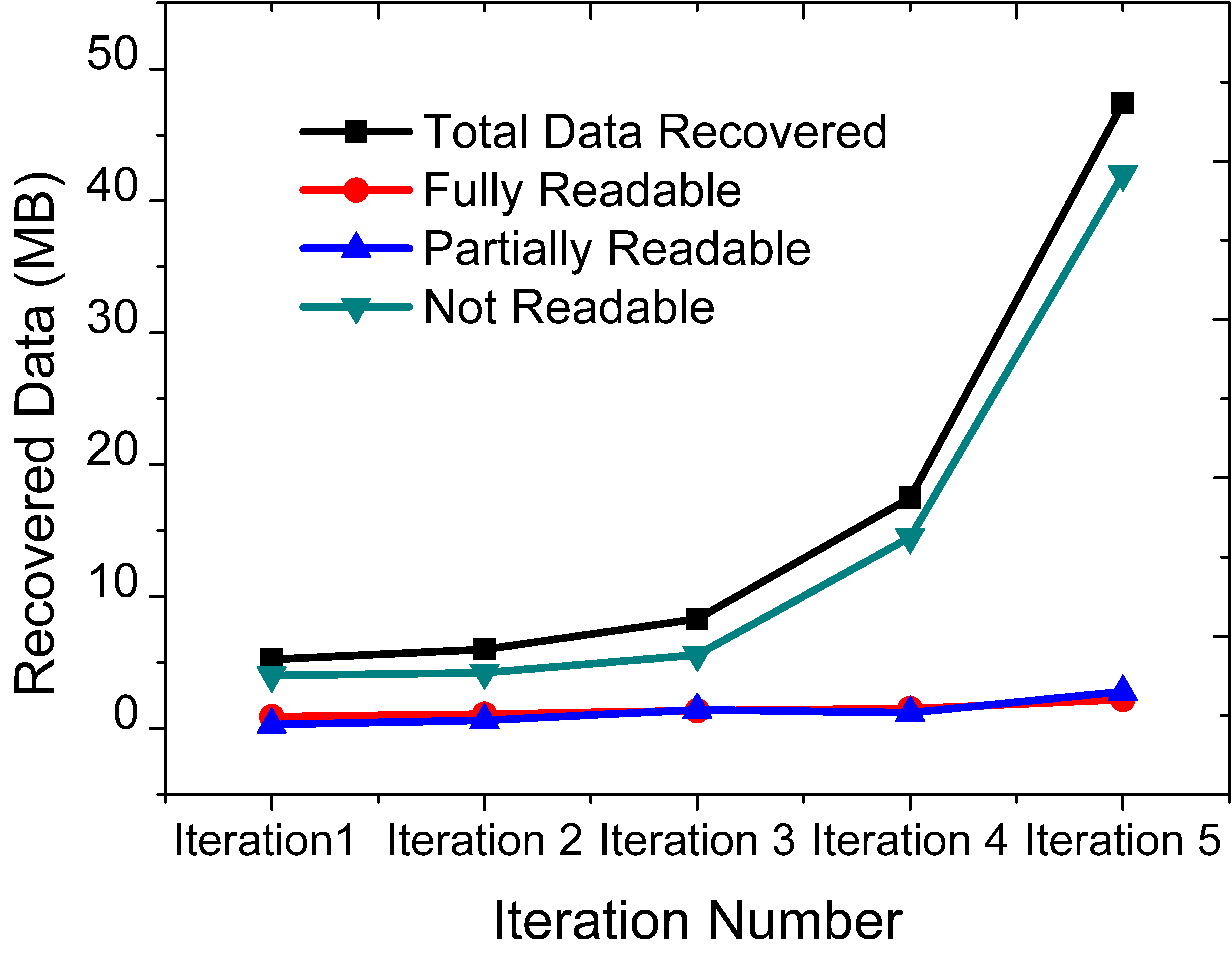}
\caption{\label{fig:rec123}Data recovery statistics over the number of iterations for NAND Flash USB when data is recovered on same storage media.}
\vspace{-5mm}
\end{figure}

\begin{figure}
\centering
\includegraphics[scale=0.45]{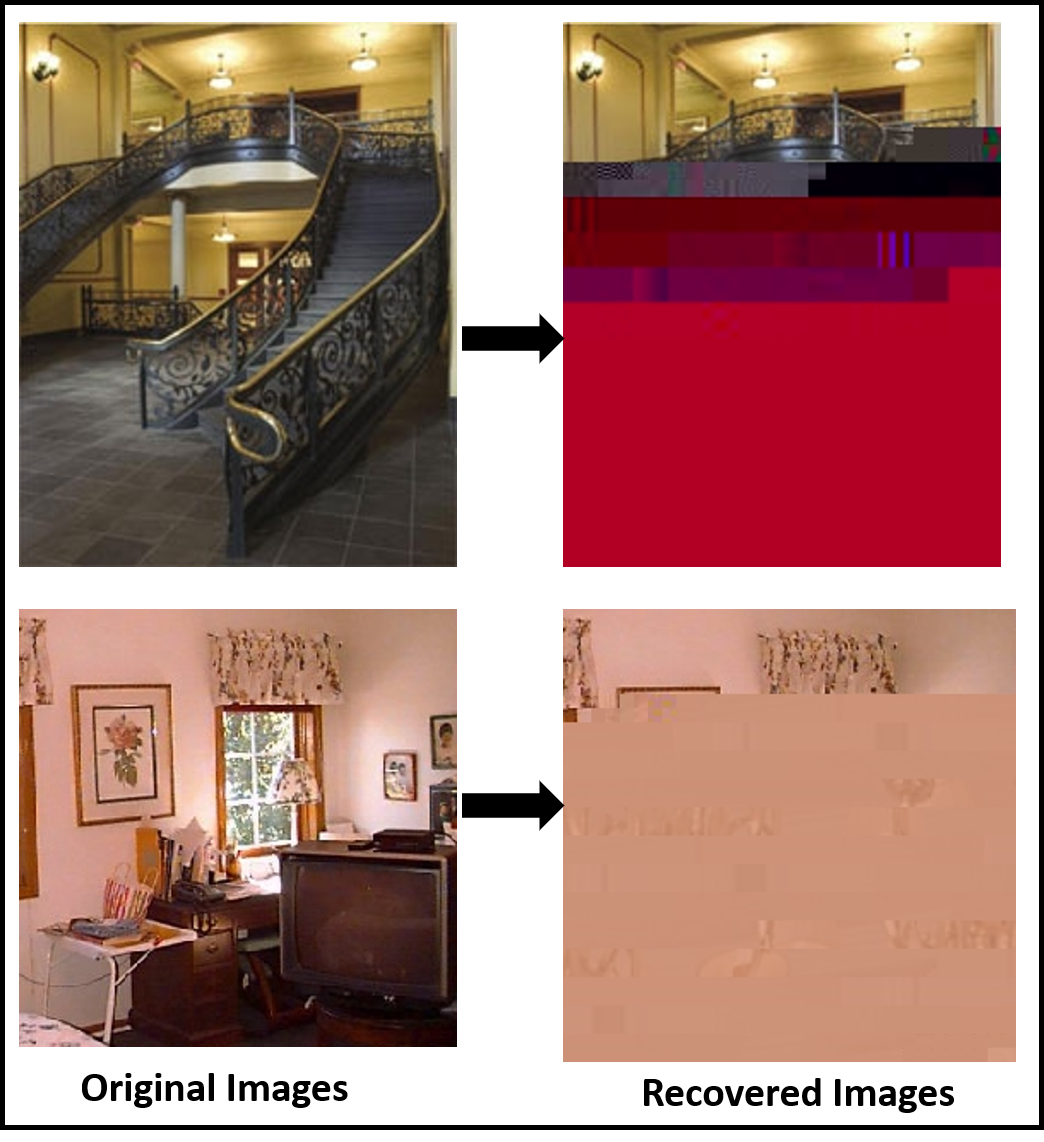}
\caption{\label{fig:partial1}Partially recovered (Distorted) sample image files when data recovery was attempted on same storage media.}
\vspace{-5mm}
\end{figure}

In these experiments the recovery tool was asked to dump the recovered data on the same USB storage media from which the recovery was being made. 
In one such experiment, first 5.25 MB of jpeg image data was written on to the Flash USB storage media. After writing into the USB media, these files were permanently deleted using Shift+Delete as claimed by the Windows OS. After which recovery software was run, 5.25 MB data was recovered on the same storage media. The recovered data was again deleted permanently and again recovery was attempted. Data was partially recovered. This has been repeated for five iterations and every time amount of recovered data got increased. The statistics of the experiment is shown in Fig.~\ref{fig:rec123}.
\\In an another experiment, 229 MB of sample text files (.pdf and .txt) and 1.5 GB of audio and video files (.mp4) were written on the storage media and then similar cyclic procedure of 'delete-recover-delete' was performed. Initially only some text files were readable (not audio and video) but over the number of iterations readable text files were reduced significantly.
\\Some key observations of the same-storage media cyclic 'delete-recover-delete' experiments were:
\begin{itemize}
\item Filenames of some files got modified or interchanged between two cycles.
\item Some files lost their original file format.
\item After repeating the experiment many times, some of the files which had lost their format in one of the previous cycles were successfully recovered with their original format.
\item Extra/Multiple copies of certain files were recovered.
\item Partial recovery of image files was also evidenced as shown in Fig.~\ref{fig:partial1}. 
\item Some files which were lost in earlier cycles were recovered in distant subsequent cycles (non-consecutive).
\item With increasing number of iterations (Fig.~\ref{fig:rec123}), size of recovered data increased and readable data decreased. 
\end{itemize}
These observations confirm the data deletion vulnerabilities in Flash based storage technology. Recovery of multiple copies of same data even after deletion shows the inverse effect of wear levelling and bad block management algorithms. Although over the number of iterations, readable data is becoming negligible but that is due to the change in operational properties of files. Data is still there and one can easily access that data by restructuring operational properties of files or in worst case by doing charge/charge contrast measurement. Data presence in non readable files can be observed by opening those files in notepad++.

\section{Conclusion}
{\label{conclusions}}
In this paper, we have presented a comprehensive study about data-deletion/sanitization related vulnerabilities in NAND Flash memory based storage technology. All the reasons giving rise to these deletion vulnerabilities at different abstraction layers are discussed in detail. Software based data recovery experiments resulted in 100 $\%$ data recovery in windows and linux based delete/Erase commands. Similar results are obtained for both NTFS and FAT filesystems. Multiple copies of same data were found during data recovery on same storage media thus confirming the inverse effect of performance enhancement techniques like wear leveling and bad block management.

\bibliographystyle{IEEEtran}
\bibliography{IEEEabrv,main.bib}

\end{document}